\documentclass[aps, prx, twocolumn, superscriptaddress]{revtex4-2}

\usepackage{amsmath,amsfonts,bm,amssymb,mathtools}
\usepackage{graphicx}
\usepackage{float}
\usepackage{subfigure}
\usepackage{color}
\usepackage{algorithm}
\usepackage{algpseudocode}
\usepackage[colorlinks=true,citecolor=blue,linkcolor=blue,urlcolor=blue]{hyperref}
\usepackage{ntheorem}
\usepackage{array}
\usepackage{multirow}
\usepackage{makecell}
\usepackage{booktabs}

\begin{document}
  
  \title{Learning Hamiltonian neural Koopman operator and { simultaneously sustaining  and discovering} conservation laws}
  \author{Jingdong Zhang}
  \affiliation{School of Mathematical Sciences, SCMS, and SCAM, Fudan University, Shanghai 200433, China}
  \affiliation{Research Institute of Intelligent Complex Systems, Fudan University, Shanghai 200433, China}
  
  \author{Qunxi Zhu}\email{qxzhu16@fudan.edu.cn}
  \affiliation{Research Institute of Intelligent Complex Systems, Fudan University, Shanghai 200433, China}
  \affiliation{Shanghai Artificial Intelligence Laboratory, Shanghai 200232, China}
  \affiliation{MOE Frontiers Center for Brain Science and State Key Laboratory of Medical Neurobiology, Fudan University, Shanghai 200032, China}
  
  \author{Wei Lin}\email{wlin@fudan.edu.cn}
  \affiliation{School of Mathematical Sciences, SCMS, and SCAM, Fudan University, Shanghai 200433, China}
  \affiliation{Research Institute of Intelligent Complex Systems, Fudan University, Shanghai 200433, China}
  \affiliation{Shanghai Artificial Intelligence Laboratory, Shanghai 200232, China}
  \affiliation{MOE Frontiers Center for Brain Science and State Key Laboratory of Medical Neurobiology, Fudan University, Shanghai 200032, China}
  
  \date{\today}
  \begin{abstract}
    Accurately finding and predicting dynamics based on the observational data with noise perturbations is of paramount significance but still a major challenge presently.
    Here, for the Hamiltonian mechanics, we propose the Hamiltonian Neural  Koopman Operator (HNKO), integrating the knowledge of mathematical physics  in learning the Koopman operator, and making it automatically {sustain and even discover} the conservation laws.
    We demonstrate the outperformance of the HNKO and {\color{black}its extension} using a number of representative physical systems {\color{black}even with hundreds or thousands of freedoms}.
    Our results suggest that feeding the prior knowledge of the underlying system and the mathematical theory appropriately to the learning framework can reinforce the capability of machine learning in solving physical problems.
  \end{abstract}
  \maketitle
  
  Accurate reconstruction of nonlinear dynamical systems solely based on the observational data with noise perturbations is a focal challenge in many fields of physics and engineering. The neural networks (NNs) equipped with the induced biases have remarkable abilities in learning and generalizing the intrinsic kinetics of the underlying systems from the noisy data, such as the Hamiltonian NNs \citep{greydanus2019hamiltonian,han2021adaptable}, the Lagrangian NNs \citep{cranmer2020lagrangian}, the neural differential equations \cite{chen2018neural, zhu2021neural, zhu2022neural}, and the physics-informed NNs \citep{raissi2019physics,karniadakis2021physics,de2021hyperpinn,lu2021physics,chen2021physics,eivazi2022physics,cai2022physics,mao2020physics,misyris2020physics}.
  These frameworks have been applied successfully to many tasks (e.g., the generative tasks \citep{toth2019hamiltonian}, the dynamics reconstruction \citep{zhong2019symplectic, sanchez2019hamiltonian, zhu2019detecting, zhu2023leveraging}, the intelligent control problems \cite{NEURIPS2022_3b91129c, zhang2023sync}, and the tipping point  detection  ~\citep{rolnick2022tackling, li2023tipping}), sharing the common idea in design--utilization of an appropriate loss function enforcing the model to nearly obey the physical principles.  Although progresses have been outstandingly achieved, these frameworks, which either enlarge the network complexity or overfit the noisy data during the training stage to decrease the loss, are suffered from the poor generalization abilities.  Recently, endowing NNs with natural physics priors becomes an effective approach to promote the sample efficiency,  robustness, and generalization ability of NNs \citep{zhang2018machine,liu2021machine,lezcano2019cheap,lezcano2019trivializations,zhang2022hamiltonian}.
  
  Recent advances in the Koopman operator theory pave a direct way to identify intrinsic kinetics using infinite-dimensional linear representations for strongly nonlinear systems \citep{brunton2021modern,mezic2005spectral,rowley2009spectral,lusch2018deep,brunton2020machine,otto2017universal}.  Several algorithms using observational data have been developed for approximating such an operator, including the dynamic mode decomposition (DMD) \citep{schmid2010dynamic,tu2013dynamic,rowley2009spectral} and the extending dynamic mode decomposition (EDMD) \citep{williams2015data,brunton2019notes}.  Although all these algorithms try to obtain parsimonious models and maintain accurate reconstructions of the unknown systems, either conservation properties or the accurate prediction cannot be obtained surely.
  
  In this Letter, inspired by the advances of physics-informed learning and the Koopman operator theory, we articulate a framework to efficiently and robustly learn the Hamiltonian dynamics based solely on the observational data even with noise perturbations.   Noticing the unitary property of the Koopman operator for the Hamiltonian dynamics, we use an orthogonal neural network to approximate the Koopman operator, so that the learned operator naturally sustains the conservation laws.   Also, we include an auto-encoder structure in the NNs to identify the nonlinear coordinate transformation, mapping the original data to a low-dimensional manifold on a sphere.  We test the proposed HNKO framework on a group of representative Hamiltonian systems and show its advantages over many mainstream methods in several aspects including robust preservation of the conservation laws and accurate prediction of the dynamical behaviors.
  
  \begin{figure}[t]
    \vskip 0.03in
    \centering\subfigure{\label{fig_sketcha}}\subfigure{\label{fig_sketchb}}
    \includegraphics[width=8cm]{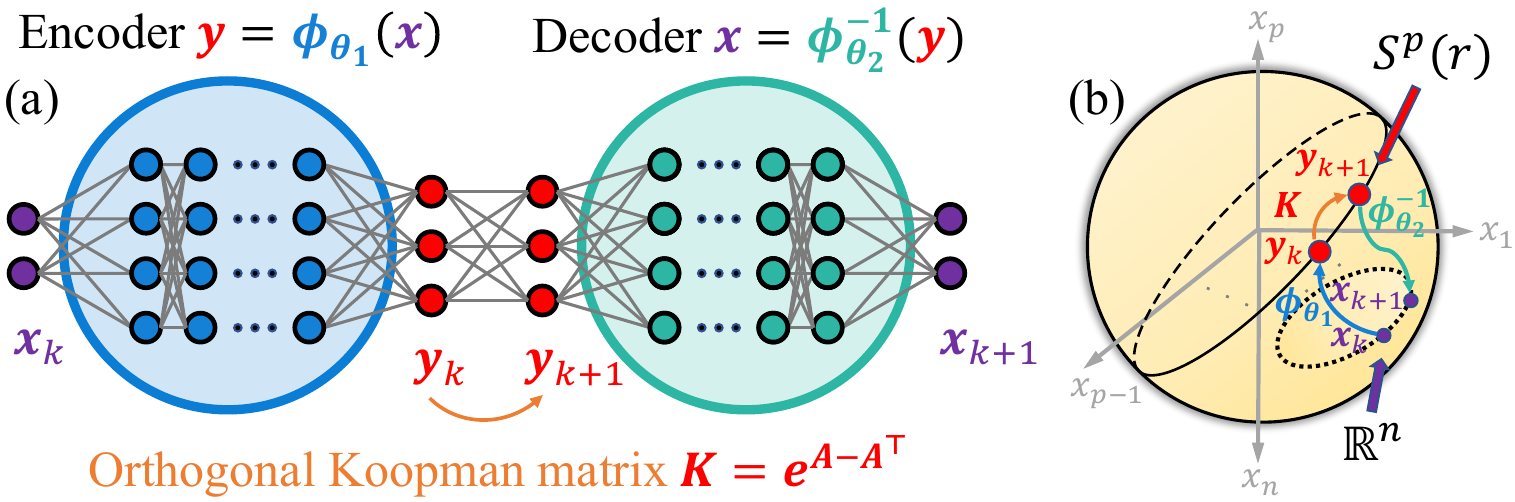}
    \caption{A sketch for the HNKO framework. (a) A combination of the auto-encoder with the orthogonal Koopman matrix $\bm{K}$ using NNs. (b) Geometrically, the encoder embeds the original data in $\mathbb{R}^n$ to some low-dimensional manifold on $p$-dimensional sphere, and the decoder reverses this process, where $\bm{K}$ maps the trajectory on the embedded manifold.}
    \label{fig_sketch}
    \vskip -0.2in
  \end{figure}
  
  \textit{Method.}---To begin with, we consider a dynamical system whose state vector $\bm{x}=(x_1,\cdots,x_n)^\top\in\mathcal{M}\subset\mathbb{R}^n$ evolves along some smooth {\it symplectic} vector field $f(\bm{x})$.  The flow mappings produced by this Hamiltonian  dynamical system form an operator group, denoted by $\{F_t:F_t(\bm{x})=\bm{x}+\int_0^tf(\bm{x}(s))\mathrm{d}s,\bm{x}\in\mathcal{M},t>0\}$. The Koopman operator $\mathcal{K}_t$ with regard to any flow $F_t$ is an infinite-dimensional linear operator, acting on the function space $\mathcal{F}=\{g:\mathcal{M}\to\mathbb{R}\}$ and satisfying $\mathcal{K}_tg=g\circ F_t$ for $g\in\mathcal{F}$. Specifically, if the time interval $\Delta t$ and the initial state $\bm{x}_0\in\mathcal{M}$ are given,  a state trajectory is generated by this flow, denoted as $\{\bm{x}_k:\bm{x}_k=F_{k\Delta t}\bm{x}_0\}_{k=0}^m$. Thus, $\mathcal{K}_{\Delta t}g(\bm{x}_k)=g(F_{\Delta t}(\bm{x}_k))=g(\bm{x}_{k+1})$.  For a sake of simplicity, we use $\mathcal{K}$ for $\mathcal{K}_{\Delta t}$ if $\Delta t$ is given.  Since, in practice, the observational data $\{\bm{x}_i\}_{{i=0}}^m$ often contain noise perturbations,  it is really difficult for the existing methods to achieve a robust and accurate approximation of $\mathcal{K}$ and simultaneously preserve the energy-like quantity in the considered Hamiltonian dynamics.  It thus motivates us to find a framework owning all these capabilities.
  
  The DMD algorithm \citep{schmid2010dynamic,tu2013dynamic,rowley2009spectral} constructs two data matrices $\bm X$ and $\bm X'$ from the observational data by $\bm X=(\bm{x}_0,\bm{x}_2,\cdots,\bm{x}_{m-1})$ and $\bm X'=(\bm{x}_1,\bm{x}_2,\cdots,\bm{x}_m)$ for $\bm X,\bm X'\in\mathbb{R}^{n\times m}$. Then, according to \citep{schmid2010dynamic}, the optimal linear operator $\bm K=\bm X'\bm X^{+}\in\mathbb{R}^{n\times n}$  satisfying $\bm K\bm X\approx\bm X'$ is regarded as an approximation for $\mathcal{K}$, where $\bm X^{+}=(\bm X^\top \bm X)^{-1}\bm X^\top$ is the Moore-Penrose pseudo-inverse. Particularly, since the Koopman operator for the conservative Hamiltonian dynamics is unitary, we should restrict the candidate $\bm K$ in the special orthogonal group $\textbf{SO}(n)=\{\bm B\in\mathbb{R}^{n\times n}~|~\bm B\bm B^\top=\bm{I},~\text{det}(\bm B)=1\}$. Then, the DMD actually solves  the vanilla optimal problem: $	\arg\min_{\bm K\in \textbf{O}(n)} \big\| \bm K\bm X-\bm X'\big\|_{\rm F}$, where $\Vert\cdot\Vert_{\rm F}$ is the Frobenius norm.   The vanilla surrogate of the DMD has two major weaknesses:
  (i) The size of $\bm K$ is limited by the system's dimension $n$, which is not large enough to approximate the intrinsically infinite-dimensional operator $\mathcal{K}$, and
  (ii) the orthogonal transformation preserves the norm of the state and hence induces its dynamics $\{\bm K^i\bm{x}_0\}_{{i=0}}^m$ embedded on a sphere, while the conserved orbit of the original Hamiltonian dynamics may not be on some $n$-dimensional sphere.
  
  To overcome the first weakness, the EDMD was developed in \citep{williams2015data,brunton2019notes} to  lift the dimension of $\bm K$ by introducing a dictionary of nonlinear observational functions $\{g_i\}_{i=1}^p$ and obtaining the augmented state $\bm{y}=(g_1(\bm{x}),g_2(\bm{x}),\cdots,g_p(\bm{x}))^\top\in\mathbb{R}^p,~p>n$. Analogous to the DMD, the two data matrices are constructed as $\bm Y=(\bm{y}_0,\cdots,\bm{y}_{m-1})$ and $\bm Y^{'}=(\bm{y}_1,\cdots,\bm{y}_m)$, which gives an approximated Koopman operator as  $\bm K=\bm Y'\bm Y^{+}\in\mathbb{R}^{p\times p}$.
  However, the EDMD uses the dictionary $\{g_i\}_{i=1}^p$ as a basis of $\mathcal{F}$ and, in practice, it requires an extremely large dictionary to approximate the coordinate function, which makes the accurate approximation inefficient and even impossible.  To reduce the computational cost and promote the representation ability, we adopt an auto-encoder NN to encode $\bm{y}=(g_1(\bm{x}),\cdots,g_p(\bm{x}))^\top$ as $\bm{y}=\phi_{{\bm{\theta}}_1}(\bm{x})$, and decode the coordinates as $\bm{x}=\phi^{-1}_{{\bm{\theta}}_2}(\bm{y})$, as shown in Fig.~\ref{fig_sketcha}. We train the weights ${\bm{\theta}}=({\bm{\theta}}_1,{\bm{\theta}}_2)$ in this auto-encoder using the loss function as $\mathcal{L}_\text{dict}({\bm{\theta}})=\sum_{{i=0}}^m\big\|\bm{x}_i-\phi^{-1}_{{\bm{\theta}}_2}(\phi_{{\bm{\theta}}_1}(\bm{x}_i))\big\|^2$ with the data.
  
  To conquer the second weakness, we in this work embed the augmented state $\bm{y}$ to some higher-dimensional sphere, denoted by $S^p(r)=\{\bm{y}\in\mathbb{R}^p~|~\Vert\bm{y}\Vert^2=r^2\}$, and obtain a constrained optimization problem as 
$\arg\min_{\bm K} \big\| \bm K\bm Y-\bm Y'\big\|_{\rm F}$ such that $\bm K\in \textbf{SO}\left(p\right)$, 
      $r^2\ge\max_{0\le i\le m}\Vert\bm{x}_i\Vert^2$,
and $\bm{y}_i=\phi_{{\bm{\theta}}_1}\left(\bm{x}_i\right)\in S^p\left(r\right),~0\le i\le m$.
  Actually, it is still difficult to solve this optimization problem because of its highly nonconvex property induced by the orthogonality constraint.
  
  To solve this problem efficiently, we first notice a fact that the Lie exponent map $\bm A\mapsto\exp(\bm A)=\bm{I}+\bm A+\frac{1}{2}\bm A^2+\cdots$ from the skew-symmetric group $\mathfrak{so}(n)=\{\bm A\in\mathbb{R}^{n\times n}:\bm A+\bm A^\top={\bm{0}}\}$ to $\textbf{SO}(n)$ is surjective \cite{lezcano2019cheap}. There exists an isomorphism $\alpha$ from $\mathbb{R}^{n(n-1)/2}$ to $\mathfrak{so}(n)$ as $\alpha(\bm A)=\bm A-\bm A^\top$ where $\bm A\in\mathbb{R}^{n(n-1)/2}$ is identified as an upper triangular matrix with the zero diagonal elements. Thus, in our framework, we represent the  orthogonal Koopman operator approximately as a parameterized form $\bm K=\exp(\alpha(\bm A))$, where $\bm A$ owns $n(n-1)/2$ learnable parameters.
  Hence, we train $\bm{K}$ using the loss function as $\mathcal{L}_\text{koop}(\bm K)=\sum_{{i=0}}^{m-1}\Vert \bm K\bm{y}_i-\bm{y}_{i+1}\Vert^2=\sum_{{i=0}}^{m-1}\Vert \exp(\alpha(\bm A))\phi_{{\bm{\theta}}_1}(\bm{x}_i)-\phi_{{\bm{\theta}}_1}(\bm{x}_{i+1})\Vert^2.$
  Significantly, our framework does not require those uninterpretable   regularization term in the loss function while all the conventional methods always use it \citep{greydanus2019hamiltonian,cranmer2020lagrangian}, and, indeed, the above configurations of mathematical physics automatically guarantee the orthogonality of the operator $\bm K$ during its training procedure.
  
  To further ensure the augmented state trajectory $\{\bm{y}_i\}_{{i=0}}^m$  on some $p$-dimensional sphere, we set $r$, the radius of the embedded sphere, as a learnable parameter and simply set the distance to the origin as another loss function $\mathcal{L}_\text{sphere}({\bm{\theta}},r)=\sum_{{i=0}}^{m}\big(\Vert \bm{y}_i\Vert^2-r^2\big)^2=\sum_{{i=0}}^{m}\big(\Vert \phi_{{\bm{\theta}}_1}(\bm{x}_i)\Vert^2-r^2\big)^2$.  We prove that the trajectory generated by $\bm K\in\textbf{SO}(p)$ belongs to a manifold of dimension at most $\lfloor p/2\rfloor$ [see Supplementary Information (SI)]. Hence, the freedom degree of the trajectory $\{\bm{y}_i\}_{i=0}^m$ is lower than $\lfloor p/2\rfloor$.
  Notice that, once the hyperplane equation $\langle {\bm{v}},\bm{y} \rangle =0$ is satisfied for any nonzero vector ${\bm{v}}$, the freedom degree for $\bm{y}$ decreases by one order. Thus, to restrict the freedom degree of the augmented states, we introduce a loss as: $\mathcal{L}_\text{deg}(q)=\sum_{k=1}^q\sum_{{i=0}}^m {\big\langle \frac{{\bm{v}}_k}{\Vert{\bm{v}}_k\Vert}, \bm{y}_i\big\rangle}^2
  = \sum_{k=1}^q\sum_{{i=0}}^m {\big\langle \frac{{\bm{v}}_k}{\Vert{\bm{v}}_k\Vert}, \phi_{{\bm{\theta}}_1}(\bm{x}_i)\big\rangle}^2$,
  where $\bm{V}=({\bm{v}}_1,\cdots,{\bm{v}}_{q})$ are learnable parameters with with ${\bm{v}}_{k}\in\mathbb{R}^p$, $k=1,2,\cdots,q\le p-2$, and $q\in \mathbb{Z}^{+}$. To guarantee the linear independence of the column vectors in $\bm{V}$, we introduce an orthogonal regularization term as $\mathcal{L}_\text{ind}=\sum_{k\neq j} {\left\langle{\bm{v}}_k, {\bm{v}}_j \right\rangle}^2$.
  
  Finally, we train the parameters $\{{\bm{\theta}},\bm{K},r,\bm{V}\}$ in a delicately-designed framework of NNs, integrating the prior knowledge of mathematical physics and using a comprehensive loss function as:
  $\mathcal{L}({\bm{\theta}},\bm{K},r,\bm{V})=\mathcal{L}_\text{dict}({\bm{\theta}})+\mathcal{L}_\text{koop}(\bm{K})+\mathcal{L}_\text{sphere}({\bm{\theta}},r)+\mathcal{L}_\text{deg}(q,{\bm{\theta}},\bm{V})+\mathcal{L}_\text{ind}(\bm{V}\textit{}),$
  where $\mathcal{L}_\text{sphere}$ with $\mathcal{L}_\text{deg}$ embeds the original $n$-dimensional data to the $(p-q-1)$-dimensional manifold on $S^p(r)$, and the adjustable hyperparameter $q$ satisfies $p-\lfloor p/2\rfloor-1\le q\le p-2$. {Significantly, the extracted features $\{g_i(\bm{x})\}_{i=1}^{p}$ from the encoder span the Koopman invariant subspace. The eigenvector $\bm{c}=(c_1,\cdots,c_p)^\top$ associated with the eigenvalue $1$ of the orthogonal Koopman matirx $\bm{K}$ induces the analytical Hamiltonian of the original system, $g_{\bm{c}}(\bm{x})=\sum_{i=1}^{p}c_ig_i(\bm{x})$ (see SI).}
  
  {\color{black}\textit{Scalability for high-dimensional systems.}--- To reduce the computational complexity of the Lie exponent operation when applying the current HNKO to any high-dimensional system, we approximate the $p$-dimensional Koopman matrix $\bm{K}$ via $\bm{K}_1\otimes\bm{K}_2$, where $(\bm{K}_i)_{p_i\times p_i}$~$(i=1,2)$ are the two orthogonal matrices with $p_1p_2=p$, implying the orthogonality of $\bm{K}$ \cite{LOAN200085}, and $\otimes$ is the Kronecker product.  We name this extension as HNKO$^\ast$. As such,  by applying the HNKO$^\ast$ with $p_{1,2}=\sqrt{p}$, we reduce the computational cost of the learnable parameters from $\mathcal{O}(p^2)$ to $\mathcal{O}(p)$ (See Appendix for more illustrations).}
  
  \begin{figure}[t]
    \hspace{-0.7cm}\centering
    \subfigure{\label{fig_threea}}
    \subfigure{\label{fig_threeb}}
    \subfigure{\label{fig_threec}}
    \subfigure{\label{fig_threed}}
    \subfigure{\label{fig_threee}}
    \subfigure{\label{fig_threef}}
    \subfigure{\label{fig_threeg}}
    \subfigure{\label{fig_threeh}}
    \subfigure{\label{fig_threei}}
    \subfigure{\label{fig_threej}}
    \subfigure{\label{fig_threek}}
    \includegraphics[width=9cm]{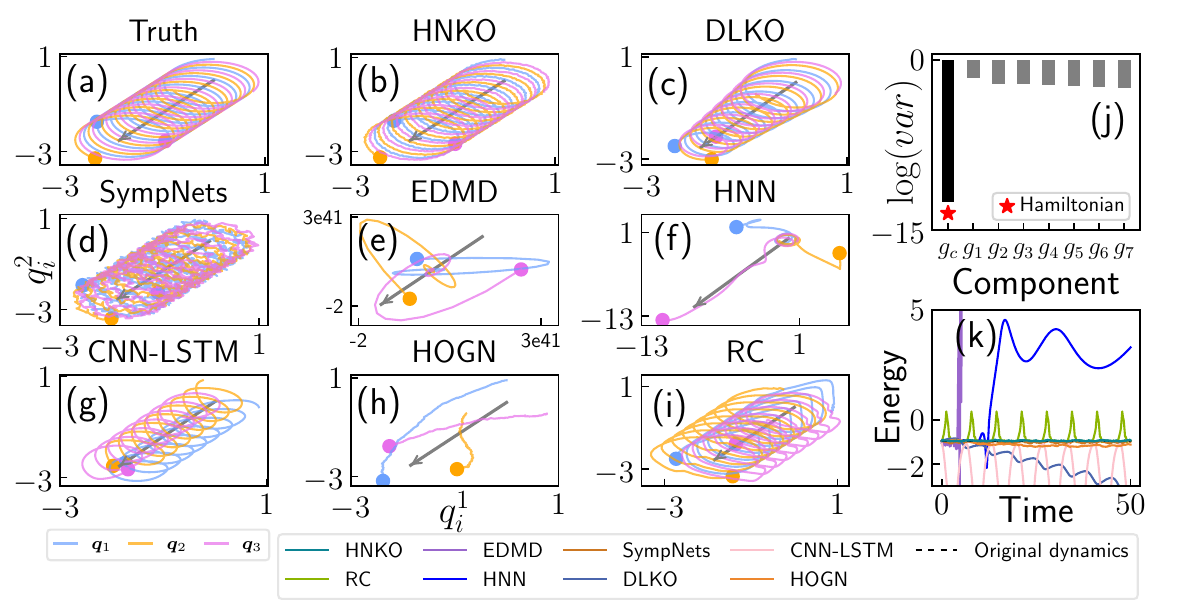}
    \vskip -0.15in
    \caption{Comparison studies on the three-body problem. (a) The  original and noise-free dynamics for the interacted bodies.  Here, the motion $\bm{q}_{i}=(q_{i}^{1},q_{i}^{2})$, and the trajectories are the projections from the original spatiotemporal space  $q_{i}^{1}$-$q_{i}^{2}$-$t$ to the phase plane with a normal vector as $(\sin(-\frac{\pi}{50})\cos\frac{\pi}{4},\sin(-\frac{\pi}{50})\sin\frac{\pi}{4},\cos(-\frac{\pi}{50}))$.
    The grey direct line indicates the time direction and the terminal positions of the three bodies are highlighted by blue, purple, and orange colors, respectively.   The reconstructed and the predicted dynamics using  {  the HNKO (b) and the other the most advanced machine learning techniques (c)-(i) are shown, respectively.  (j) The temporal variance in the logarithm scale ($\log(var)$) of the features and the discovered Hamiltonian.} (k) The system's energies using different methods change over the time.  Here, we set $m_{1,2,3}=g=1$.}
    \label{fig_three}
    \vskip -0.2in
  \end{figure}
  
  Next, we numerically show several advantages of our framework, and validate the natural existence of the conservation in our operator.  This makes the framework extremely suitable for dealing with the noise-perturbed data produced by the Hamiltonian dynamics.   Indeed, we take a number of representative nonlinear Hamiltonian systems of physical significance, including the many-body problem, the  stiff spring oscillator, and the Korteweg-De Vries (KdV) equation.  Throughout, the noise-perturbed data are set as $\{\tilde{\bm{x}}_i=\bm{x}_i+\bm{\xi}_i\}_{{i=0}}^m$, where the information of the used noise $\{\bm{\xi}_i\}$ is provided in SI.
  
  %\noindent\textit{HNKO for many-body problem.}
  \textit{Many-body problem.}---We consider the classic $n$-body problem.   First, we focus on the case of $n=3$, where the canonical Hamiltonian dynamics \citep{valtonen2006three}: $\dot{{\bm{q}}_i}=H_{{\bm{p}}_i}$, $\dot{{\bm{p}}_i}=-H_{{\bm{q}}_i}$ with ${\bm{p}}_i,{\bm{q}}_i \in\mathbb{R}^2~(i=1,2,3)$ representing the space coordinates and the momenta, respectively.  Here, $H=\sum_{i=1}^3\frac{m_i}{2}\Vert{\bm{p}}_i\Vert_2^2-g\sum_{i< j}m_im_j(\Vert{\bm{q}}_i-{\bm{q}}_j\Vert_2)^{-1}$ is the total conserved energy with mass $m_i$ and gravitational constant $g$. The three bodies interact with the others through an attractive force from gravity, and the force tends to infinity when the two particles get close to each other.  As shown in Fig.~\ref{fig_threea}, the orbits of the three bodies evolve in a higher dimensional spatiotemporal space, which results in a challenge for accurate prediction based on the noise-perturbed observational data.
  
  We show the prediction performance using the HNKO framework.  Particularly, we train the NNs with the noise-perturbed data on the time interval $[0,5]$, less than a period, and produce the trajectories using the trained model on the interval $[0,50]$.   As shown in Fig.~\ref{fig_threeb}, the reconstructed and the predicted dynamics are preserved in high fidelity and for a fairly long time~\footnote{The predicted trajectories by the HNKO framework seem to be a bit jagged, which reflects a peculiar ability of our framework in sustaining the feature of noise perturbations.  In fact, using the technique of embedding the original data to the manifold on $S^p(r)$ not only extracts maximally the conservation law rooted in the original data, but also withstands the noise effect efficiently.
   All these indicate the necessity and importance of integrating the priors of mathematical physics in reconstructing and predicting a system.}.  We also test the methods of the EDMD with a dictionary of at most $2^{\text{nd}}$-order Hermite polynomials \citep{williams2015data}, the HNN \citep{greydanus2019hamiltonian}, the SympNets \citep{jin2020sympnets}, the deep learning Koopman operator (DLKO) \citep{lusch2018deep}, {the CNN-LSTM~\cite{mutegeki2020cnn,vidal2020gold}, the Hamiltonian ODE graph networks (HOGN)~\cite{sanchez2019hamiltonian}, and the reservoir computing (RC)~\citep{pathak2018model,RC_pre}.} on the same generated noise-perturbed data.  The produced trajectories [see Figs.~\ref{fig_threec}-\ref{fig_threei}] {\color{black}either cannot sustain the conservation laws or} diverge extremely fast.   {Figure~\ref{fig_threej} displays a successful discovery of the Hamiltonian using the HNKO.}  Additional comparisons in Fig.~\ref{fig_threek} suggest that the HNKO framework attains the least prediction errors and only has the ability of maintaining the conservation law.  In SI, we further show successful examples using our framework when the length of the training data within a period is even shorter, and also show the reconstruction ability of the HNKO in dealing with the chaotic dynamics of the three-body system.

  \begin{figure}[t]
    \centering
    \subfigure{\label{fig_keplera}}
    \subfigure{\label{fig_keplerb}}
    \subfigure{\label{fig_keplerc}}
    \subfigure{\label{fig_keplerd}}
    \subfigure{\label{fig_keplere}}
    \subfigure{\label{fig_keplerf}}
    \includegraphics[width=8.6cm]{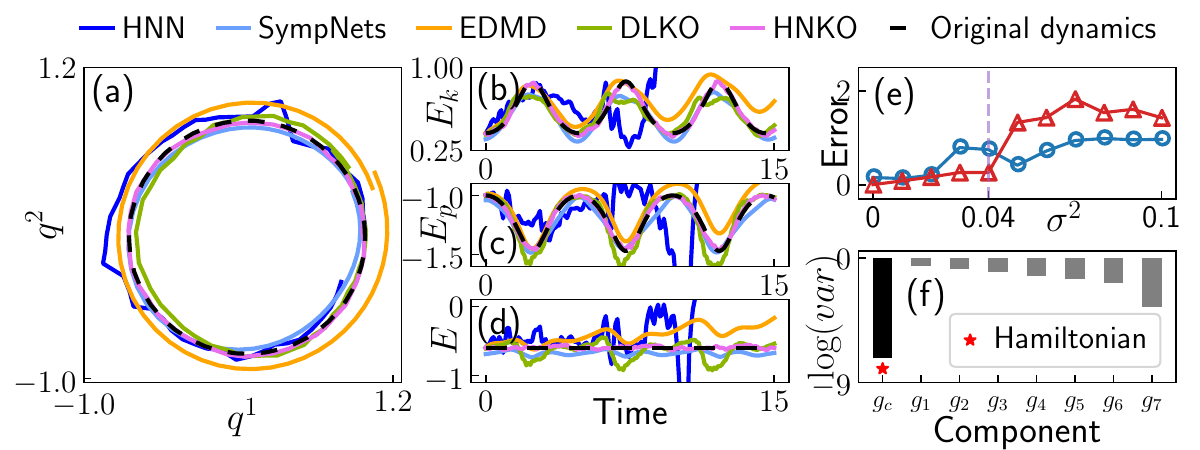}
    \caption{Comparison studies on the Kepler problem. (a) The original dynamics and the predicted phase orbits using different methods and the coordinate ${\bm{q}}=(q^{1},q^{2})$.  The changes of the kinetic energy $E_k$ (b), the potential energy $E_p$ (c), and the total energy $E$ (d) over the time for different methods.  (e) Prediction errors of the state and the energy change with the noise variance $\sigma^{2}$, using HNKO.  Here,  $m=g=1$.
    {(f) $\log(var)$ of the features and the discovered Hamiltonian.}}
    \label{fig_kepler}
    \vskip -0.3in
  \end{figure}
  
  Next, we test our framework for a specific case: $n=1$, the classic Kepler problem as: $\dot{{\bm{q}}}=H_{{\bm{p}}}$, $\dot{{\bm{p}}}=H_{{\bm{q}}}$ with $H=\frac{m}{2}\Vert{\bm{p}}\Vert_2^2- gm^2\Vert{\bm{q}}\Vert_2^{-1}$~\citep{arnol2013mathematical}.    We also compare this framework with the other methods.  As shown in Figs.~\ref{fig_keplera}-\ref{fig_keplerd}, the HNKO framework obtains the best prediction result and maintains the conservative law perfectly, while the corresponding results obtained by the EDMD and the HNN are far away from the original dynamics and energies.   Notice that, for this case, the results by the EDMD seem to be better than those in Fig.~\ref{fig_threec} obtained for the above case $n=3$.  This suggests that the EDMD method using a $2^{\rm nd}$-order dictionary could provide acceptable prediction results for complex systems of lower dimensions.  However, due to the curse of dimensionality,  it cannot directly use a dictionary of higher-order polynomials to deal with the data produced from even a three-body problem, which thus impacts its practical usefulness. {\color{black}In SI, we further show the efficacy of the HNKO framework in reconstruction and prediction for the chaotic $3$-body case and for the $n$-body problem with $n\gg3$ as well.  Particularly, to show the advantages of the HNKO over the existing methods, we reconstruct the periodic solution, previously-found in \citep{fusco2011platonic}, of the $24$-body problem in the $3$-dimensional space (see Appendix Tab.~\ref{table2}).}  In addition, the robustness of the HNKO against the noise perturbation is demonstrated from the view of state and energy predictions, as shown in  Fig.~\ref{fig_keplere}. {Significantly, Figure~\ref{fig_keplerf} shows a compelling and successful identification of the Hamiltonian using the HNKO.}

\begin{figure}[t]
    \centering
    \subfigure{\label{fig_stiffa}}
    \subfigure{\label{fig_stiffb}}  \subfigure{\label{fig_stiffc}} \subfigure{\label{fig_stiffd}}
    \includegraphics[width=8.3cm]{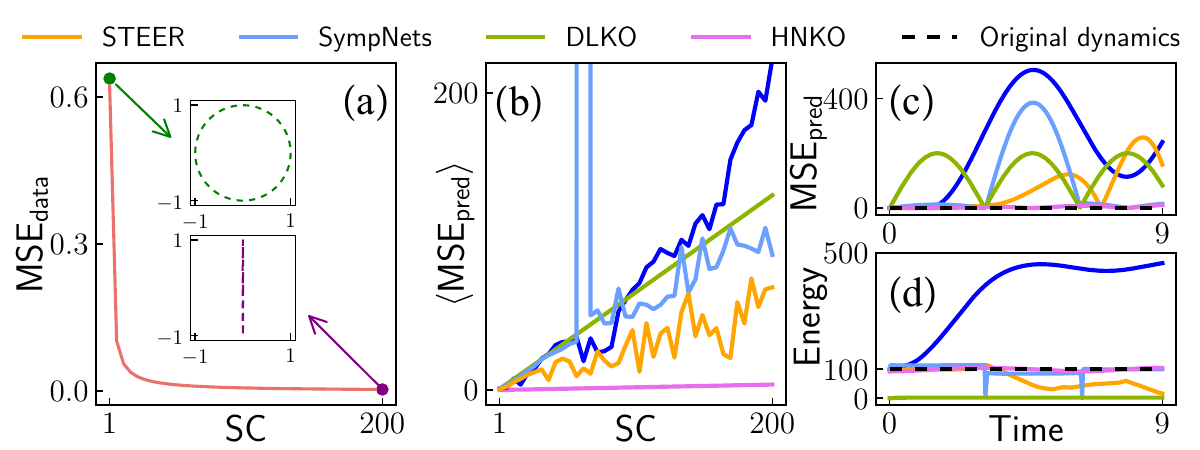}
    \caption{Comparison studies on the stiff mass-spring system.
    (a) The mean square error (MSE$_{\text{data}}$) between the normalized trajectories  and the vertical line segment under different SCs. The subfigures show the trajectories in the least and the stiffest cases. (b) The mean prediction MSE on time interval $[0,9]$ of different methods over SCs. The prediction MSE (c) and the energy (d) over the time for different methods.
    }
    \label{fig_stiff}
  \end{figure}

%  \noindent\textit{HNKO for stiff mass-spring system.}
  \textit{Stiff mass-spring system.}---We consider the friction-free mass-spring system: $\dot{q}=p/m,~\dot{p}=-kq$, where $(q,p)\in\mathbb{R}^2$ are the canonical coordinates representing the position and the momentum, $m$ is the mass, and $k$ is the elastic coefficient~\citep{fowles1999analytical}. With large $k$ and $m$, the system becomes a typical slow-fast system with the stiffness
  coefficient (SC) as $\sqrt{km}$ ~\citep{kim2021stiff}.
  The system's conservative total energy $\frac{1}{2}kq^2+\frac{1}{2m}p^2$ corresponds to the elliptic phase orbits with an eccentricity going to $1$ as the SC goes to infinity.  In Fig.~\ref{fig_stiffa}, under a high SC, the trajectory looks like noise series wandering along the line segment, which leads to a failure of reconstruction and prediction using the ODE solver-based learners \citep{kim2021stiff}.
  
  % However, the HNKO can alleviate the effects of stiffness because the encoder embeds the original stiff data into non-stiff data on a high dimensional sphere (see Fig.~S4 in SI).
  We compare the  prediction performance of HNKO under different SCs, with existing methods including the STEER, a better method for solving stiff ODEs problems~\citep{ghosh2020steer}. Figures~\ref{fig_stiffb}-\ref{fig_stiffd} shows the  HNKO alleviates the effects of stiffness while the other methods perform unstably and diverge fast as SC grows. 
 % However, the HNKO can alleviate the effects of stiffness because the encoder embeds the original stiff data into non-stiff data on a high dimensional sphere (see Fig.~S4 in SI).
 %  Spefically, we show the  prediction performance of HNKO under different SCs, and verify its advantages over the HNN, {\color{black}the SympNets, the DLKO}, and the STEER, a better method for solving stiff ODEs problems~\citep{ghosh2020steer}. Figure~\ref{fig_stiffb} shows the long time prediction ability of HNKO robust against the stiffness, while the other methods perform unstably and diverge fast as SC grows. Figures~\ref{fig_stiffc}-\ref{fig_stiffd} present the best performances of the HNKO on sustaining both prediction accuracy and energy law for the stiffest case of SC=$200$.
  
  \begin{figure}[t]
    \centering
    \subfigure{\label{fig_kdva}}
    \subfigure{\label{fig_kdvb}}
    \subfigure{\label{fig_kdvc}}
    \subfigure{\label{fig_kdvd}}
    \subfigure{\label{fig_kdve}}
    \subfigure{\label{fig_kdvf}}
    \subfigure{\label{fig_kdvg}}
    \subfigure{\label{fig_kdvh}}
    \includegraphics[width=8.5cm]{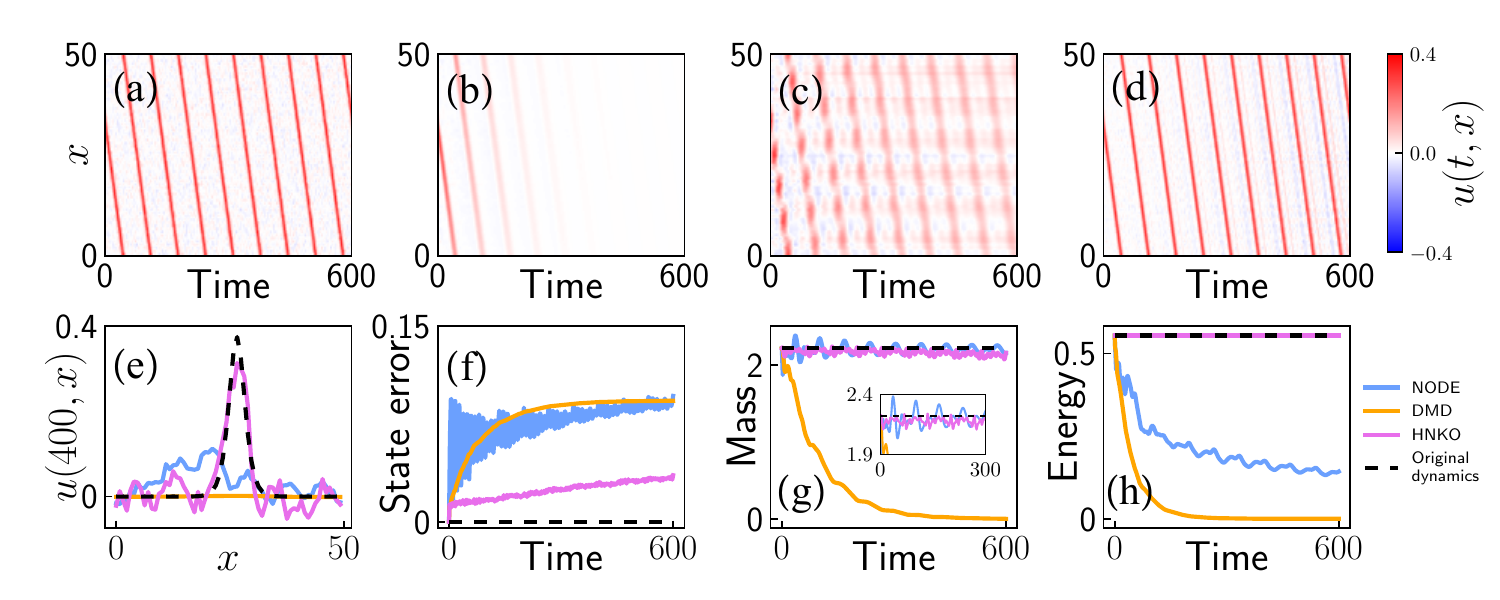}
    \caption{Comparison studies on the KdV equation. (a) The dynamics produced by the KdV equation and perturbed with the Gaussian noise $\mathcal{N}(0,0.03\bm{I})$. The predicted trajectories using the DMD (b), the NODE (c), and the HNKO (d). Different solitary waves produced by using different methods at $t=400$ (e), the prediction error for the state (f),  the total mass (g), and the total energy (h) over the time using different methods.
    The inset in (g) zoom in the changes of mass using the NODE and the HNKO.  Here, we introduce the discretization to ${\bm{u}}(t,\cdot)$ on $[0,50]$ by $64$ predefined grid points. }
    \label{fig_kdv}
    \vskip -0.25in
  \end{figure}
  
%  \textit{HNKO for KdV equation.}
  \textit{KdV equation.}---We finally consider the KdV equation ${\bm{u}}_t+{\bm{u}}_{xxx}-6{\bm{u}}{\bm{u}}_x=0$, a Hamiltonian partial differential equation with infinitely many integrals of motion (IOM)~\citep{miura1968korteweg}.
  This equation is used to describe the behavior of shallow water waves with periodic solitary wave phenomenon~\citep{zabusky1971shallow}, as shown in Fig.~\ref{fig_kdva}. Two low-order IOMs of the system--$\int {\bm{u}}\mathrm{d}x$ and $\int {\bm{u}}^2\mathrm{d}x$, corresponding to the mass and the energy conservation--are selected as the indexes to evaluate the prediction performance.  
  
  % Since the KdV equation cannot be formulated in a canonical form with space coordinates and momenta, the symplectic structure-based methods including the HNN are not applicable to prediction task. The EDMD is not applicable because $n^{64}$, the size of the $n^{\rm th}$-order Hermite polynomials dictionary, incurs unbearably computational cost.
  
  We conduct comparison studies using respective methods: the HNKO, the DMD, and the neural ordinary differential equation (NODE), a recently-developed and widely-used framework~\citep{chen2018neural}. The HNN and EDMD cannot work here due to \footnote{Since the KdV equation cannot be formulated in a canonical form with space coordinates and momenta, the symplectic structure-based methods including the HNN are not applicable to prediction task. The EDMD is not applicable because $n^{64}$, the size of the $n^{\rm th}$-order Hermite polynomials dictionary, incurs unbearably computational cost.}.
  % We train the NNs with the noise-perturbed data on the time interval $[0,360]$ and produce the trajectories using the trained model on the interval $[0,600]$. 
  As shown in Figs.~\ref{fig_kdva}-\ref{fig_kdvd}, the HNKO robustly keeps the periodicity and the consistency of solitary wave solutions for a long time, while the DMD only holds a short-term forecast, showing a rapid decay and the trained NODE shows high fluctuations due to its less robustness against noise interference.  After a sufficiently long time evolution, the required solitary wave behavior is only observed in the trained model using the HNKO, as shown in Fig.~\ref{fig_kdve}.  More significantly, due to the orthogonality rooted in the HNKO,  the conservation laws of both mass and energy are sustained only in the trained model using the HNKO [see Figs.~\ref{fig_kdvf}-\ref{fig_kdvh}].
    {\color{black} In  Appendix Tab.~\ref{table1},
we further successfully apply the HNKO$^\ast$ to cope effectively with the high dimensional tasks with the number of the grid points even as $1024$.  }

  \textit{Concluding remarks.}---We have proposed a machine learning framework to approximate the Koopman operator for the Hamiltonian dynamics.  Different from the mainstream NN methods \citep{greydanus2019hamiltonian, cranmer2020lagrangian, kaiser2018discovering}, our framework integrates typical mathematical physical structures of orthogonality and flexibility and thus has natural advantages in accurately reconstructing the Hamiltonian dynamics from the noise perturbed data, and achieving accurate prediction simultaneously and perfectly with energy conservation~\citep{liang2021stiffness,chen2019symplectic}.
  Therefore, it can be applied to reconstruction and prediction problems in the real-world scenarios where the conservation laws are persistent but the collected data are more or less shuffled and contaminated. { More importantly, based on the Koopman invariant space spanned by the features in the encoder and the orthogonal spectral decomposition of the Koopman matrix, we show that our HNKO owns an ability of discovering the conservation laws with analytical expression, which makes an essential step towards solving the critical problem of discovering conservation laws~\citep{liu2021machine,lu2023discovering,iten2020discovering,kaiser2018discovering}. }

 {\color{black}
 
 {\it Acknowledgements.}--Q. Zhu is supported by the China Postdoctoral Science Foundation (No. 2022M720817), by the Shanghai Postdoctoral Excellence Program (No. 2021091), and by the STCSM (Nos. 21511100200, 22ZR1407300, and 23YF1402500). W. Lin is supported by the NSFC (No. 11925103) and by the STCSM (Nos. 22JC1402500, 22JC1401402, and 2021SHZDZX0103).  The computational work presented in this Letter is supported by the CFFF platform of Fudan University. 
 
{\it Appendix on reducing computational cost by Kronecker product}-- Since the Kronecker product of two orthogonal matrices is still an orthogonal matrix \citep{LOAN200085}, one can reduce the computational complexity for computing the $p$-dimensional orthogonal matrix $\bm{K}$ by replacing it with the Kronecker product of two low order orthogonal matrices $(\bm{K}_1)_{p_1\times p_1}$ and $(\bm{K}_2)_{p_2\times p_2}$. Originally, obtaining the orthogonal matrix $\bm{K}$ involved calculating the Lie exponent $\exp(\bm{A})$ of a skew-symmetric matrix $\bm{A}$, which results in a computational complexity of $\mathcal{O}(p^3)$ due to the high power of $\bm{A}$ in the expansion. However, with the Kronecker method, the computational cost can be reduced to $\mathcal{O}(p_1^3+p_2^3)$. Since $p_1p_2=p$ and it can be easily verified that the minimum of the function $p_1^3+{p^3}/{p_1^3}$ is $2p^{{3}/{2}}$. Thus, the Kronecker product operation can reduce the original computational cost to at least $\mathcal{O}(p^{{3}/{2}})$. Additionally, for extremely high dimensional tasks, the order of the computational cost can be further reduced by repeating the Kronecker operation. For instance, by designing 
$
\bm{K}=(\bm{K}_1)_{p_1\times p_1}\otimes\cdots\otimes(\bm{K}_m)_{p_m\times p_m}
$ 
in an order of $\mathcal{O}\left(\sum_{i=1}^mp_i^3\right)$ with $p_1\times \cdots \times p_m=p$ and $p_1\approx \cdots \approx p_m \approx p^{1/m}$, the computational cost of the matrix product in calculating the Lie exponent is reduced to $\mathcal{O}(m\cdot p^{{3}/{m}})$. For brevity, we only consider the most simple case of the Kronecker operation $\bm{K}=(\bm{K}_1)_{p_1\times p_1}\otimes(\bm{K}_2)_{p_2\times p_2}$, with $p_1\approx p_2 \approx \sqrt{p}$, in this work. Notably, this condition is easily satisfied by choosing $p$ as a square number.

  {\color{black}
{\it Appendix on comparison studies}--We present the comparison studies on different reconstruction and prediction tasks in Tabs.~\ref{table2} and \ref{table1}, using the HNKO, its extension HNKO$^\ast$, and other mainstream methods.
  	
\begin{table}[htb]
     \centering
     \caption{\color{black} Comparison studies of different methods on the 24-body problem using full (144-dimensional) and partial (72-dimensional) observational data.  Here,   the performance (Perf.) is shown in terms of the 2-Wasserstein distance (2WD) and the time average of the MSE, denoted by  $\langle \text{MSE}\rangle$, between the real data and the prediction data.}
\label{table2}
%		\resizebox{\linewidth}{!}{
    \begin{tabular}{lccccccc}
				\toprule
				\toprule
				\multicolumn{1}{c}{\multirow{2}{*}{Data}} &\multicolumn{1}{c}{\multirow{2}{*}{Perf. $\downarrow$}} & \multicolumn{5}{c}{Model} &    \\
				\cmidrule(lr){3-8}  
				& &HNKO & DLKO & DMD & SympNets  & HNN   \\
				\midrule 
				\multirow{2}{*}{~Full}  &{$\langle\text{MSE}\rangle$}	&\textbf{3.63}	&12.89	&10.13	&1341.94	 &48.95	\\
				  &{2WD}	&\textbf{3.66}	&26.34	&9.48	&1080.75	 &49.17	\\		  
				 \cmidrule(lr){1-2}  
				\multirow{2}{*}{~Partial} &{$\langle\text{MSE}\rangle$} &\textbf{0.54}	&0.97	&0.97	&- 	&-	\\
				&{2WD} &\textbf{0.48}	&0.94	&0.98	&- 	&-	\\
				\midrule
				\bottomrule
			\end{tabular}
\end{table}
  
   \begin{table}[htb]
\centering
\caption{{\color{black}Comparison studies of different methods on the KdV equation. Here, $s$ is the uniform discretization in the space. The prediction $\text{MSE}$ and the training time (Tt) are shown, respectively, illustrating the high performance using the HNKO and its extension.}}
    \label{table1}
\resizebox{\linewidth}{!}{
\begin{tabular}{lcccccccccccc}
\toprule
\toprule
\multicolumn{1}{c}{\multirow{2}{*}{Model}}   & \multicolumn{5}{c}{$\langle\text{MSE}\rangle$ $\downarrow$} & \multicolumn{5}{c}{Tt (\# Seconds) $\downarrow$}   \\
\cmidrule(lr){2-6}  \cmidrule(lr){7-11} 
 &$s=64$ & 144 & 256 & 576 & 1024  &64 & 144 & 256 & 576 & 1024   \\
\midrule 
{DMD}   &0.52 &7.3e4 &2.93 &2.55 &1.68 &- &- &- &- &-\\
{DLKO} &10.73 &33.07 &2.22 &4.55 &8.40 &\textbf{37} &87 &254 &866 &2449\\
{NODE}   &0.52 &1.18 &14.34 &4.9e6 &1.8e8 &595 &8943 &1.4e4 &2.9e4 &1.8e5\\
{HNKO}  &\textbf{0.42}  &\textbf{0.50}  &\textbf{1.71} &2.79  &10.56  &62  &262  &646 &8280  &3.7e4 \\
{HNKO$^\ast$}  &1.09  &2.43 &6.21  &\textbf{1.07}  &\textbf{0.93} &46  &\textbf{59} &\textbf{64} &\textbf{175}  &\textbf{426}  \\
\midrule
\bottomrule
\end{tabular}
}
\label{tab2}
\end{table}
~}
  
  \bibliography{main}
%  \bibliographystyle{apsrev4-2}

 % \section{Appendix}

\end{document}